# TWO- AND THREE-QUBIT ROOM-TEMPERATURE GRAPHENE QUANTUM GATES

Daniela Dragoman[1] and Mircea Dragoman[2]

[1]Univ. Bucharest, Faculty of Physics, P.O. Box MG-11, 077125 Bucharest, Romania

[2]National Institute for Research and Development in Microtechnology (IMT), P.O. Box 38-160, 023573 Bucharest, Romania

**Abstract**

Proposed configurations for the implementation of graphene-based CNOT and Toffoli gates working at room temperature are presented. These two logic gates, essential for any quantum computing algorithm, involve ballistic Y junctions for qubit implementation, quantum interference for qubit interaction and oblique gates for optimizing the output, and can be fabricated using existing nanolitographical techniques. The proposed configurations of CNOT and Toffoli quantum logic gates are based on the very large mean-free-paths of carriers in graphene at room temperature.

Quantum computers promise massive parallel computation with the help of quantum gates, which process quantum bits, known as qubits [1]. The implementation, manipulation and read-out of qubits, which are a superposition of states governed by the laws of quantum mechanics, is completely different from that of the classical bits "0" and "1", implemented by transistors with two switching states. However, despite the impressive progress in the implementation of quantum gates and algorithms [2], quantum computer technology is very far from superseding classical computation, which has attained a high degree of integration, of about one billion transistors on a single silicon chip. The main difficulty faced by quantum computation is the rapid loss of coherence of quantum states, attempts to alleviate this problem including the use of atomic clock transitions [3] and quantum error correction algorithms [4].

Many solutions to overcome quantum decoherence and to assure integration and miniaturization of solid-state quantum computers have been proposed in the last years. Promising results have been obtained, for example, using the well-developed semiconductor nanotechnology, especially the silicon nanotechnology [5], and in superconductors [4], where quantum circuits containing thousands of elements on a chip were fabricated. Spin qubits, in particular, have received a lot of attention [6-9]. Even if the decoherence time is now as large as a few seconds, allowing tens of thousands of qubit operations to be performed [2,6], the present-day solid-state quantum computing systems work at extreme low temperatures (not to mention other requirements on pressure, magnetic fields, etc.), such that a commercially available quantum computer is still far away.

New hope has emerged with the discovery of graphene, in which charge carriers obey unique physical principles. Proposals of quantum computation based on spin qubits in graphene nanoribbons [10], graphene quantum dots [11], and graphene antidot lattices [12] require extremely small temperatures and/or specific carving of the graphene sheet, while valley-based qubits [13] require particular magnetic field configurations.

In this paper we propose an implementation of room-temperature two- and three-qubit logic gates in graphene based on the ballistic transport regime in this material, without the requirement of additional magnetic fields. The "0" and "1" quantum states are encoded using Y-branch switches [14,15], and the coupling between qubits is optimized using oblique electrostatic gates. We have demonstrated recently that ballistic graphene field effect transistors using oblique electrostatic gates can be fabricated on 60-70% of the surface of a 4 inch wafer chip covered with graphene monolayer grown by CVD [16], which demonstrates that the technology to fabricate our proposed graphene-based quantum gates shows relatively high reproducibility on large surfaces. Thus, circuits containing successions of quantum graphene gates could be fabricated on such areas, with minimum nanolithography features of about 10-20 nm and even less, the room-temperature ballistic transport regime in high-quality graphene monolayers being preserved for mean-free paths of up to 400 nm if graphene is deposited on $SiO_2$, or up to 1 $\mu$m in graphene deposited over boron nitride [17]. Very recently, it was demonstrated that these already impressive mean-free paths at room temperatures could attain even larger values, of more than 10 $\mu$m [18], for 20-100 nm wide graphene nanoribbons epitaxially grown on SiC. Such huge room-temperature mean-free paths are the main advantage of graphene over any other materials, which could eventually attain comparable mean-free path values only at very low temperatures. This advantage implies that quantum circuits based on ballistic transport in graphene could be easily fabricated using standard nanolithography methods, such as electron-beam lithography.

The schematic configuration of a two-qubit CNOT graphene gate is represented in Fig. 1. It consists of two nanoribbons, corresponding to the two entries: the control qubit $C$ and the input qubit denoted by *in* in Fig. 1, which interact/interfere in a region of length $L$ and width $W$ when the control qubit is in the logic state $|1\rangle$. The two logic states of $C$, $|0\rangle$ and $|1\rangle$, are identified with the outputs of the Y-junction following the entry $C$, the charge carriers with the wavefunction $|1\rangle$ passing also through a region in which an electrostatic potential $V_G$ induced by an oblique gate modifies their amplitude and phase. Note that the Y-junction implements also the superposition of the logic states $|0\rangle$ and $|1\rangle$, essential for quantum computing [14,15]. The inset in Fig. 1 illustrates the dependence of the transmission coefficient $T = |t|^2$ through the gate on $V_G$ for electrons with energy $E = 0.1$ eV, an incidence angle of $15^\circ$ and a gate length $L_0 = 50$ nm.

Assuming that the width of the interference region is $W = 100$ nm, and that the wavefunction in the input nanoribbon $i$, with $i = 1,2$, has the form $\Psi_i(y; y_i, \Delta y) \propto a_i \exp[-(y - y_i)^6/(\Delta y)^6]$, we have represented in Figs. 2(a) and 2(b) the evolution of the modulus of the wavefunction in the interference region as a function of $x$ for the cases when the logical value of the control bit is $|0\rangle$ and $|1\rangle$, respectively. In these simulations $a_1 = t$, $y_1 = -W/4$ for the input corresponding to the logical value $|1\rangle$ of $C$ ($a_1 = 0$ when $C$ is $|0\rangle$) and $a_2 = 1$, $y_2 = W/4$ for the $in$ qubit and $\Delta y = W/7.5$ in both cases, $V_G = 40$ meV, the incidence angle on the oblique gate is $15^\circ$, and a phase difference between the wavefunction of the $C$ and $in$ qubits corresponding to a difference between path lengths $\Delta L = 10$ nm was considered. In calculating the wavefunction in the interference region, we have added the contribution of each mode in the interference region.

In order to implement a CNOT gate, the two outputs, $out$1 and $out$2, with definite logic states, should interchange their logic states when the control qubit is $|1\rangle$. Assuming that the output nanoribbons are identical, the fraction of the total transmission coefficient in these nanoribbons, denoted by $T_1$ and $T_2$, respectively, are shown in Fig. 3 as a function of the length of the interference region $L$. The red and blue curves represent $T_1$ and, respectively, $T_2$, while with solid and dotted lines we have illustrated the cases when the logic values of $C$ are, correspondingly, $|1\rangle$ and $|0\rangle$. The transmission fractions $T_1$ and $T_2$ (as well as the form and evolution of the total wavefunction in the interference region) depend on the voltage applied on the oblique gate and the angle of incidence. These parameters can be chosen such that $T_1$ and $T_2$ are separated as well as possible for a given $L$ value. For an unambiguous operation of the CNOT gate, the logical states of *out*1 and *out*2 can be well defined if the dotted (and solid) curves of different colors are well separated and if the dotted curves are well separated from the solid curves of the same color. This can be achieved in our CNOT configuration if $L > 3L_0 = 150$ nm. Over propagation lengths of about 150 nm the transport should be ballistic, condition that could easily be met in graphene-based devices. Under these conditions, logical values $|1\rangle$ and $|0\rangle$ (or vice-versa) can be assigned to *out*1 and *out*2, respectively, if $T_1 < 0.5$ and $T_2 > 0.5$, these two values being interchanged if $C$ is in the state $|1\rangle$.

To implement the universal three-qubit Toffoli gate, for example, we need three entries: two control qubits, $C1$ and $C2$ and an input qubit *in*, the interaction between them interchanging the logical value of in only when both $C1$ and $C2$ are in the state $|1\rangle$. A possible graphene-based implementation of such a Toffoli gate is represented in Fig. 4, Y-junctions being again used to define the logical states of the control bits and the output states with well defined logic values. In this configuration, the *in* qubit interacts in the interference region of width $W$ and length $L$ with control qubits when the logic states of $C1$ and $C2$ are $|10\rangle$, $|01\rangle$ and $|11\rangle$. We can distinguish between these cases assuming that in the first two situations $a_1 = (1/2)t$ and in the last one $a_1 = t$. Then, the evolution of the modulus of the wavefunction in the interference region when the logic states of $C1$ and $C2$ are $|00\rangle$, $|10\rangle$ or $|01\rangle$, and $|11\rangle$ are illustrated in Figs. 5(a)-(c), respectively, for the same parameters as above, except that $V_G = 45$ meV and $\Delta L = 20$ nm. The corresponding fractions of the transmitted coefficient collected by the outputs *out*1 and *out*2 are represented in Fig. 6 with red and blue curves, respectively, as a function of the length of the interference region $L$. The solid, dotted and dashed-dotted lines correspond to the cases when the logic states of $C1$ and $C2$ are $|11\rangle$, $|00\rangle$, and $|10\rangle$ or $|01\rangle$, respectively. For the Toffoli gate, unambiguous computing is achieved when the solid lines for each color are well separated from the dotted and dashed-dotted lines of the same color, condition that can be achieved for small $L$ values, $L < 1.5L_0 = 75$ nm; ballistic transport over such propagation lengths is not difficult to achieve. For the example chosen in the simulations in Fig. 6, the logical values $|1\rangle$ and $|0\rangle$ (or vice-versa) can be assigned to *out*1 and *out*2, respectively, if $T_1 < 0.3$ and $T_2 > 0.7$, respectively, these values being interchanged if both $C1$ and $C2$ are in the logical state $|1\rangle$.

In conclusion, we have demonstrated that room-temperature CNOT and Toffoli logic gates can be implemented using the ballistic transport regime in graphene. These quantum gates can then be cascaded and form quantum circuits, the decoherence being avoided as long as the total dimension of the circuit is smaller than the mean-free path, parameter which takes a very large value in graphene, even at room temperature. The proposed configurations of CNOT and Toffoli gates are based on Y junctions to achieve superpositions of quantum logic states and on interference regions to assure interactions between qubits. Oblique gates are used in both cases to optimize the outputs. The CNOT and Toffoli gates can be implemented at room temperature with the existing nanotechnologies of a standard clean room, while the input, manipulation and readout of these quantum gates involve measurements of charge flow and application of gate voltages, procedures that are well controlled experimentally.

**Figure Captions**

Fig. 1  Schematic configuration of a two-qubit CNOT graphene gate. Inset: Dependence of the transmission coefficient on the potential energy on an oblique gate.

Fig. 2  Evolution of the modulus of the wavefunction in the interference region for the cases when the logical value of the control bit is (a) $|0\rangle$ and (b) $|1\rangle$.

Fig. 3  Fraction of the total transmission coefficient in the output nanoribbons, $T_1$ (red lines) and $T_2$ (blue lines), as a function of the length of the interference region when the logic values of $C$ are $|1\rangle$ (solid lines) and $|0\rangle$ (dotted lines).

Fig. 4  Schematic representation of a graphene-based Toffoli gate.

Fig. 5  Evolution of the modulus of the wavefunction in the interference region when the logic states of $C1$ and $C2$ are (a) $|00\rangle$, (b) $|10\rangle$ or $|01\rangle$, and (c) $|11\rangle$.

Fig. 6  Fraction of the transmitted coefficient collected by the outputs $out1$ (red lines) and $out2$ (blue lines) as a function of the length of the interference region when the logic states of $C1$ and $C2$ are $|11\rangle$ (solid lines), $|00\rangle$ (dotted lines), and $|10\rangle$ or $|01\rangle$ (dashed-dotted lines).

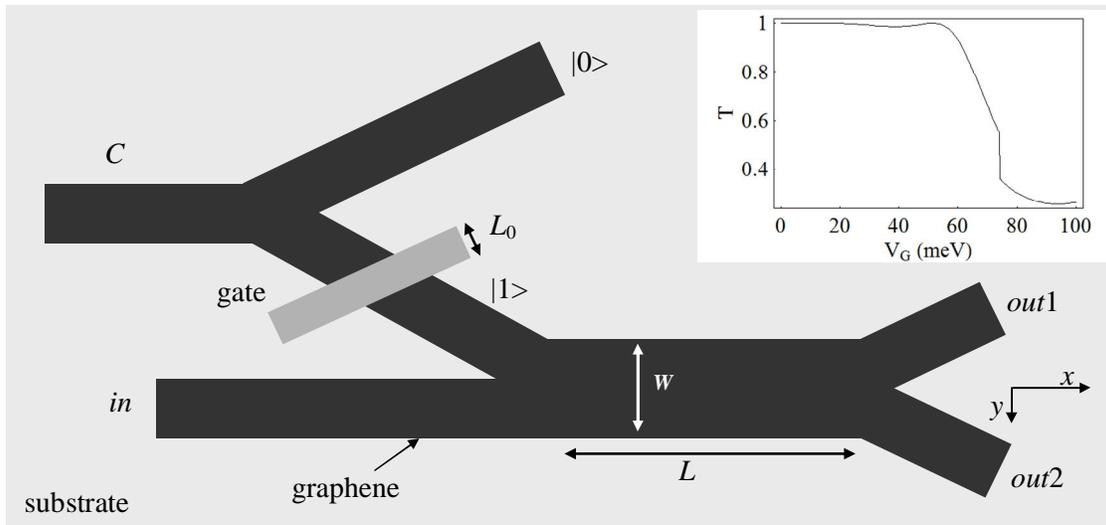

Fig. 1

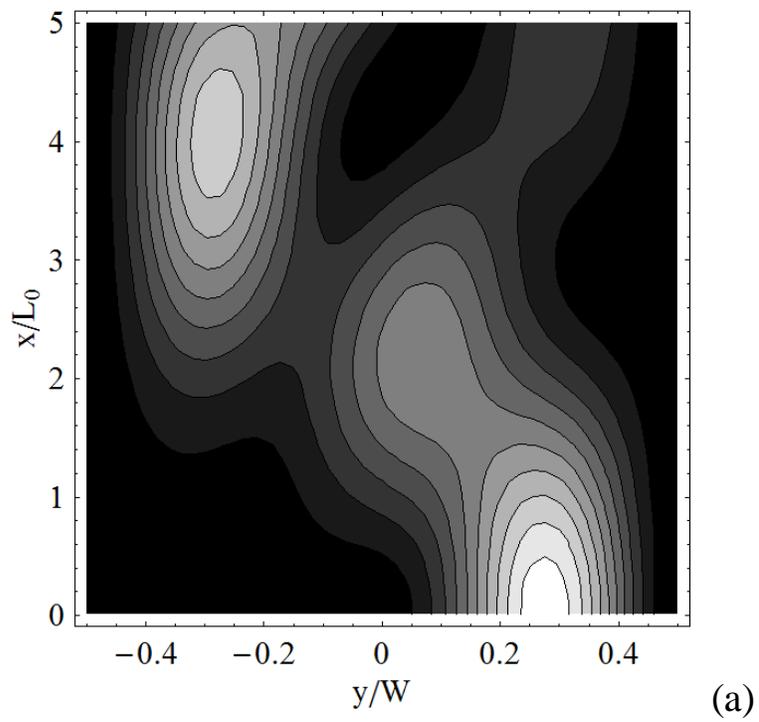

(a)

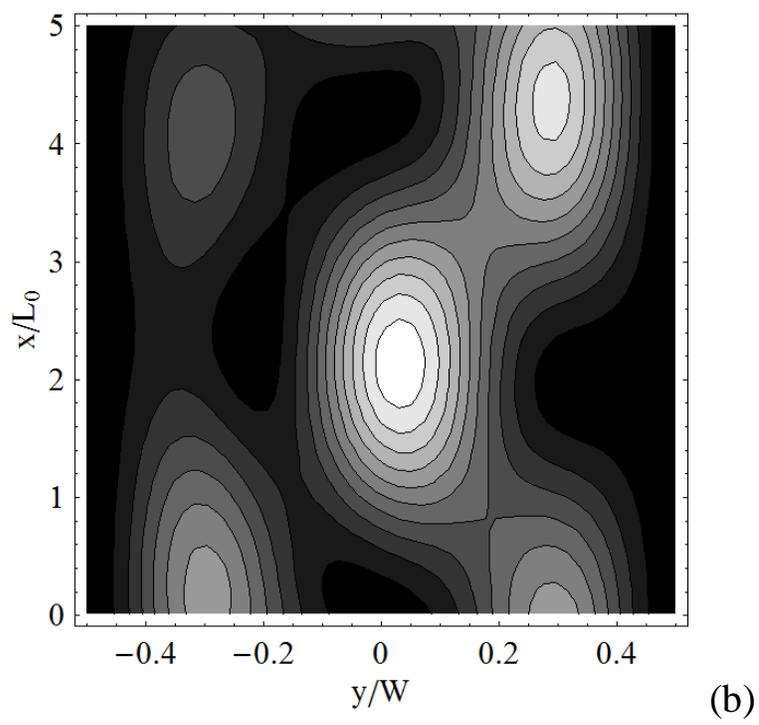

(b)

Fig. 2

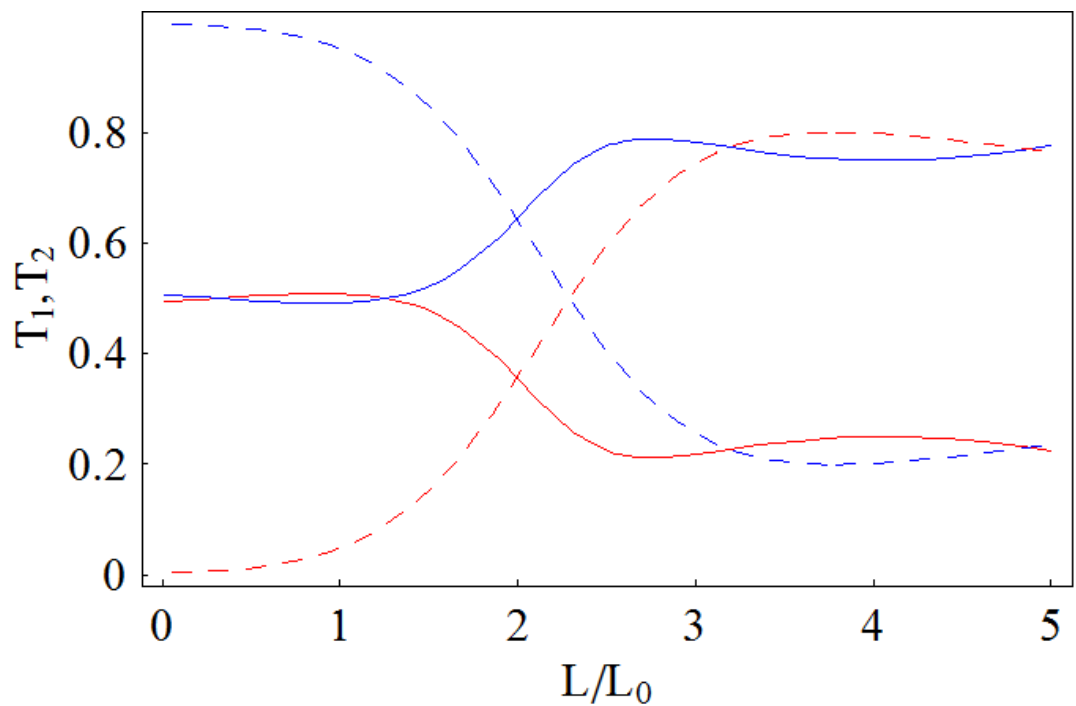

Fig. 3

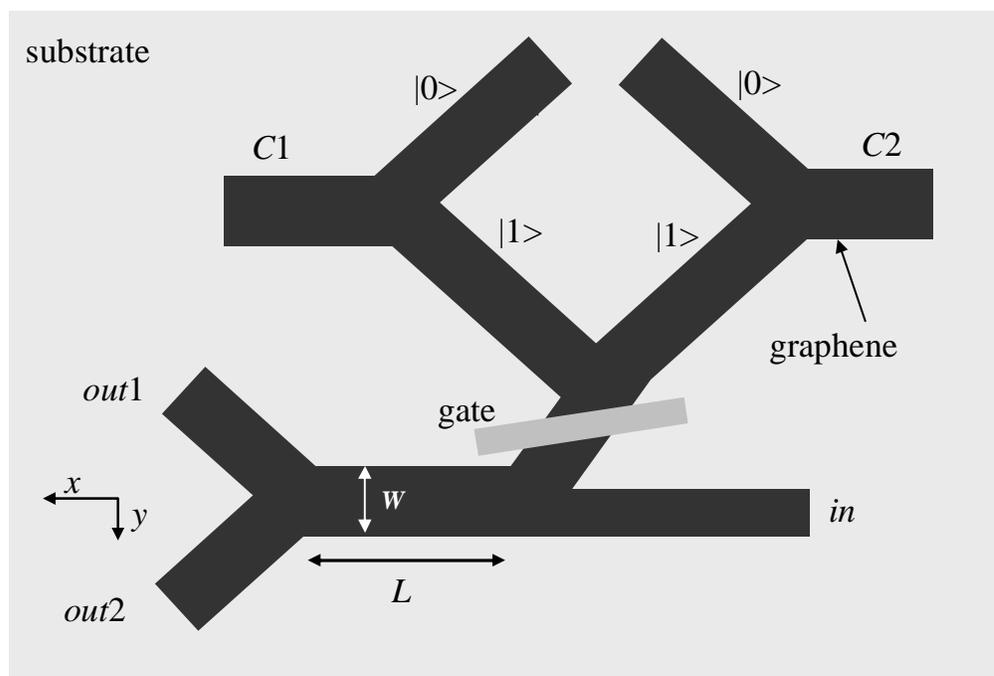

Fig. 4

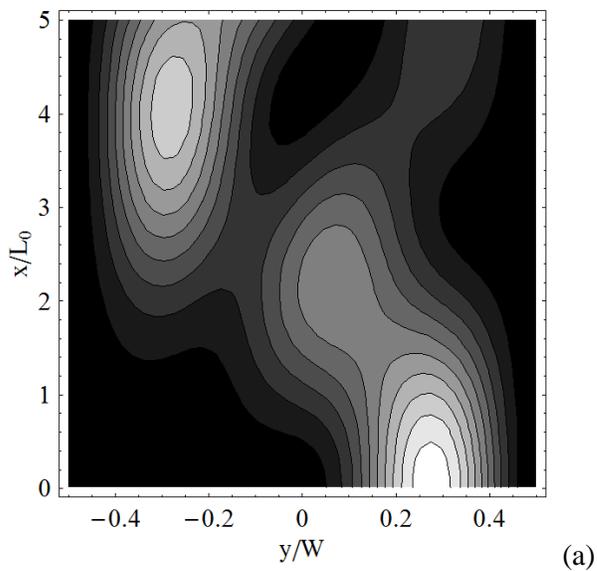

(a)

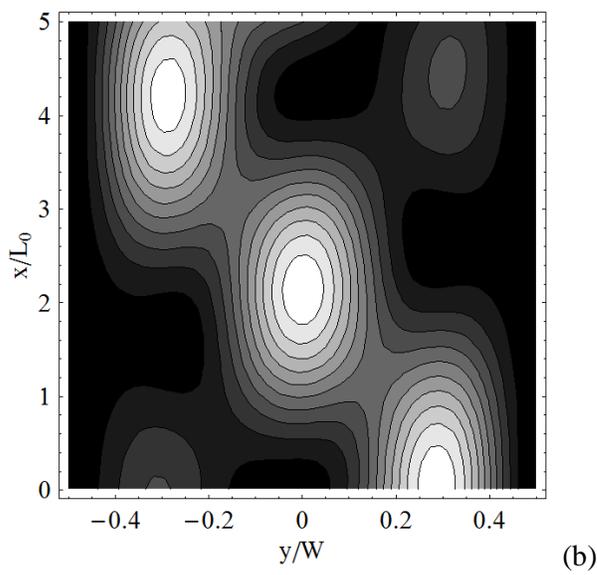

(b)

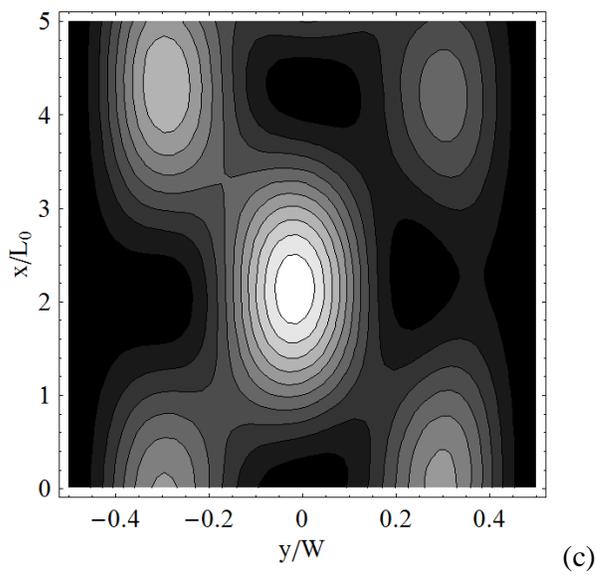

(c)

Fig. 5

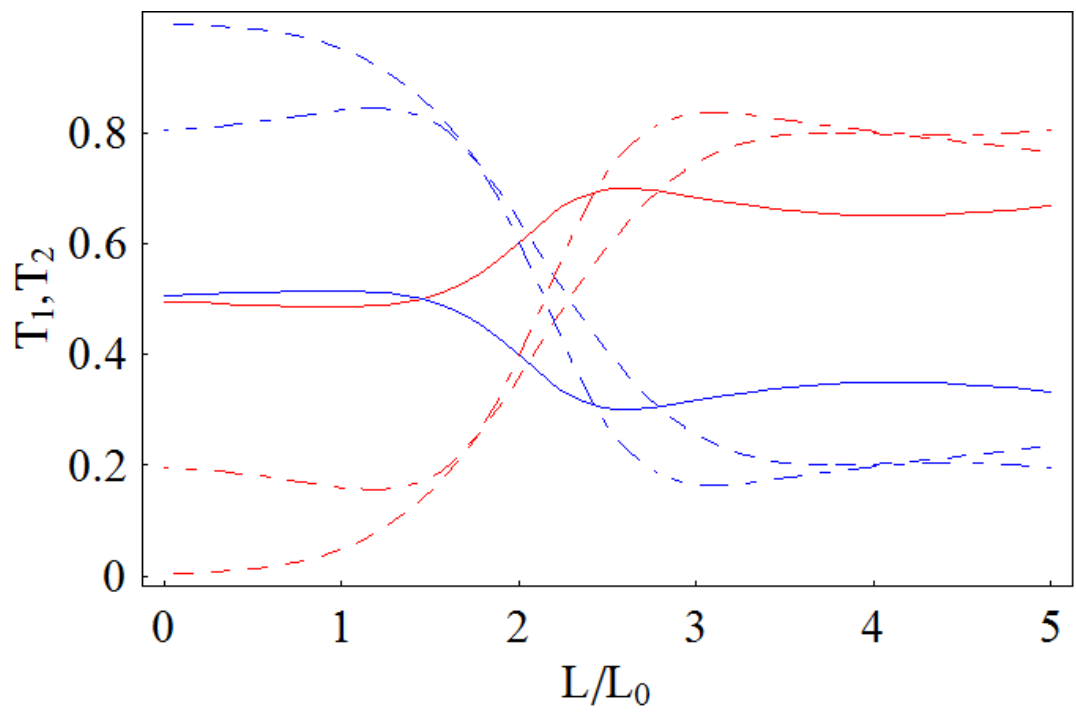